\documentclass{emulateapj}
\usepackage{graphicx}

\newcommand{\Kepler}{\emph{Kepler\ }}

\usepackage{natbib}
\usepackage{commath}

\begin{document}

\title{The Planetary Mass-Radius Relation and its Dependence on Orbital Period  as Measured by Transit Timing Variations and Radial Velocities}
\author{Sean M. Mills$^1$ and Tsevi Mazeh$^2$}
\affil{$^1$The Department of Astronomy and Astrophysics, The University of Chicago\\ 5640 S. Ellis Ave, Chicago, IL 60637, USA}
\affil{$^2$School of Physics and Astronomy, Tel Aviv University}

\begin{abstract}

The two most common techniques for measuring planetary masses---the radial velocity (RV)  and the transit timing variations (TTVs) techniques---have been observed to yield systematically different masses for planets of similar radii. Following Steffen (2016), we consider the effects of the observational biases of the two methods as a possible cause for this difference. We find that at short orbital periods ($P<11$ day), the two methods produce statistically similar results, whereas at long periods ($P>11$ day) the RV masses are systematically higher than the TTV ones. We suggest that this is consistent with an RV detection-sensitivity bias for longer periods. 
On the other hand, we do find an apparently significant difference between the short and the long-period planets, obtained by both observing techniques---the mass-radius relationship parameterized as a power law has a steeper  index at short periods than at long periods.
We also point out another anticipated observational bias between the two techniques---multiple planet systems with derived RV masses have substantially larger period ratios than the systems with TTV mass derivation.

\end{abstract}

\section{Introduction}


Several thousand exoplanets have been discovered and characterized to date. The transit method has been the most numerically successful, with the \Kepler survey alone characterizing the periods and radii of more than 4,000 planets and planet candidates \citep[e.g.,][]{2016Coughlin,2016ApJ...822...86M}. Usually, the transit light curves yield  only the radii of the transiting planets, provided we know the radii of their parent stars. However, if a star hosts multiple planets, the mutual gravitational perturbations of the planets 
may induce observable transit timing variations (TTVs), yielding constraints on the planetary masses 
\citep[e.g.,][]{2005Agol, 2005Holman, 2015ApJ...802..116D, 2016Jontof, 2016Hadden}. 
The second most prolific method of planet observation is the radial velocity (RV) technique, in which the stellar reflex motion of the parent star is measured as the planet moves in its orbit \citep[e.g.,][]{2007ARA&A..45..397U,2011arXiv1109.2497M}.
Combined with transits, this method also yields planetary masses and radii.

The many planetary masses and radii derived from the RV and TTV techniques 
enabled us to study the mass-radius (M--R) relation of exoplanets \citep[e.g.,][]{2013Weiss, 2014Weiss}, 
which is crucial for our understanding planetary formation, evolution, and structure \cite[e.g.,][]{2007Seager,2014Lopez,2016Lee}.
We use these findings to study the M--R relation below 8 $R_\mathrm{Earth}$ here. Above
this limit the planetary radius depends only weakly on the mass, because of the dominance of electron degeneracy pressure \citep[e.g.,][]{1969ApJ...158..809Z, 2007Seager, 2012ApJ...744...59S}. 

\begin{figure*}
\centerline{
\includegraphics[width=\textwidth]{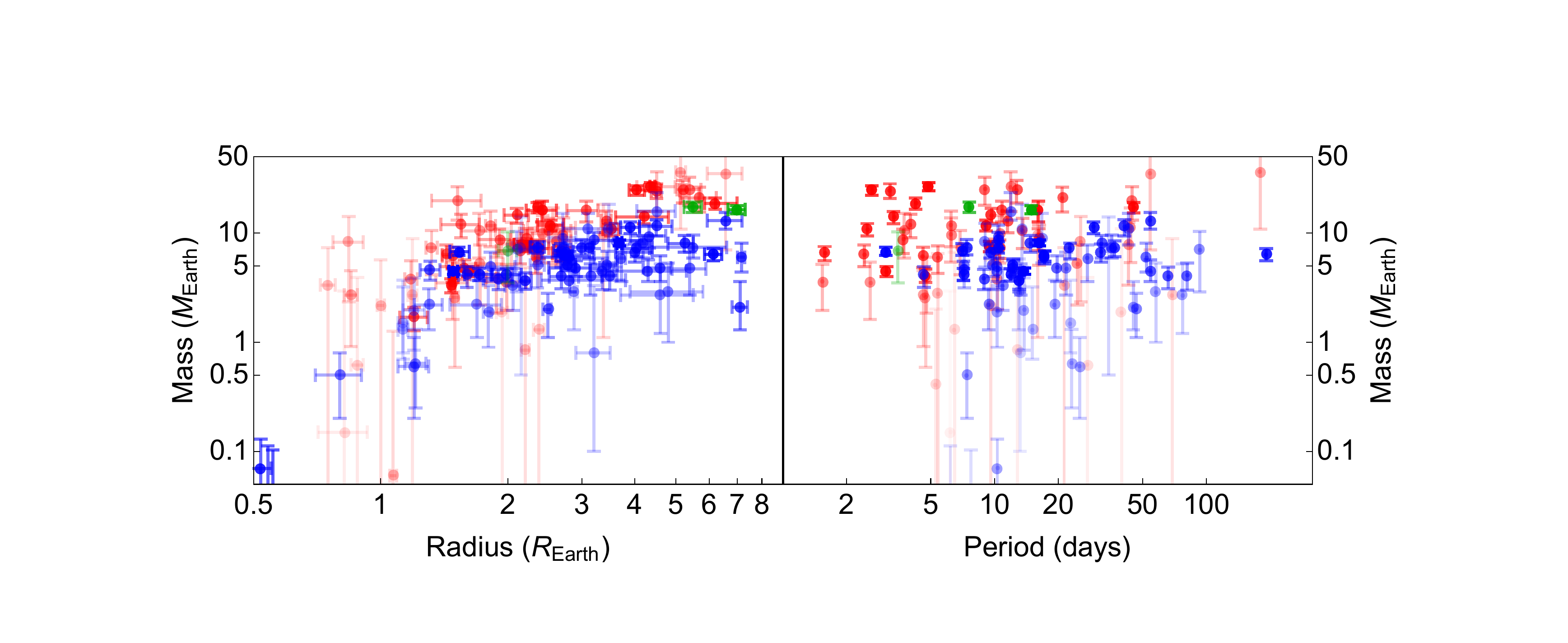}
}
\caption{Masses, radii, and orbital periods (along with 1-$\sigma$ uncertainties) for all planets as measured by RV (red), TTV (blue), or a combined RV-TTV analysis (green). The opacity of the points is decreased as the fractional uncertainties of the measurements rise so that points with large error bars do not distract the eye. Note that in both the left (M--R) panel and the right (M--P) panel, the red (RV) points generally lie above and to the left of the blue (TTV) points.
 }
\label{fig:mrpall}
\end{figure*}

However, it has been pointed out that for planets of radii less than $\sim$8 $R_\mathrm{Earth}$, planetary masses measured via RVs seem to be systematically larger than the masses measured by TTVs \citep[][henceforth S16]{2016MNRAS.457.4384S}. This observation can not be explained by the fact that the two techniques are sensitive to different radius regimes, because the difference between the two techniques holds true at any given specific radius, not merely for the distribution as a whole. This can be seen in the left panel of Figure~\ref{fig:mrpall}, an update to Figure~1 of S16, with planets color-coded by the means of their mass characterization---blue color for the TTVs, red for RVs, and green for simultaneous characterization. 

The planet data in the figure come from \cite{2016Jontof} and references therein, as well as updates from \cite{2016HaddenA, 2016ApJ...816...95G, 2016Natur.533..509M,2016ApJ...823..115D,2016ApJ...818...36P, 2016MacDonald,2017Mills}, cases where masses are robustly inferred in \cite{2016Hadden}, 
and a more complete inclusion of all positive mass planets from \cite{2014ApJS..210...20M}. We do not include measured ``negative mass'' RV planets from \cite{2014ApJS..210...20M}, which are the result of statistical fluctuations, because a similar set of ``negative mass'' statistical planets from TTV data is not available. 
We also use the default prior masses from \cite{2016Hadden} for consistency with all other measurements.

We show in Section~\ref{sec:overlap} that for most of the systems with masses derived by the two techniques, the two resulting masses are consistent with each other. Therefore the mass difference can not be explained by assuming that one technique is systematically biased.

As pointed out by S16, one basic difference between the two techniques is their sensitivity as a function of the planetary orbital period. We discuss this difference in  Section~\ref{sec:period} and show that indeed the masses and radii coming from the two techniques have different period ranges. Furthermore, there seems to be a significant difference between the M--R relation for planets at short and long orbital periods. We show that these two effects, with an assumed threshold detection for the RV techniques for long orbital period, can account for the observed difference between the RV and TTV masses. 
Section~4 points to another difference, the period ratio of the orbital periods of adjacent planets, between the systems studied by the two techniques. 
Section~\ref{sec:discussion} briefly summarizes our findings. 

\section{Observational Overlap}
\label{sec:overlap}

Several exoplanet systems have both well-measured RV and TTV signals that allow for the determination of the mass of the same planet by both techniques. It is interesting to consider these systems to see if we can detect any significant difference between the masses derived by the two techniques. We restrict the discussion here to non-zero mass measurements at the 2-$\sigma$ level published for both methods at the time of writing. In Table~\ref{table:diff} we quote the RV and the TTV masses, in Earth masses, and the RV$-$TTV mass difference in units of $\sigma$ for each planet.
Our sample totals 8 planets including one hot Jupiter \citep{2011ApJS..197....7C,2013ApJ...778..185M,2013Weiss,2015MNRAS.454.4267B, 2016ApJ...823..115D,2012MNRAS.426..739H,2015ApJ...812L..18B}. We do not include Kepler-9 \citep{2010Holman}, since there are no published RV analyses independent of TTV analysis.
We see that, excepting Kepler-89 d which differs at the 4-$\sigma$ level, the mass measurements of both methods for an individual planets are all consistent at the 1- to 2-$\sigma$ level. The mix of negative and positive difference values do not reveal any obvious systematic bias of the two methods relative to each other. A one-sample Kolmogorov-Smirnov Test of the offsets suggests that the distribution is not distinguishable from a Normal distribution---as one would expect if the two techniques are unbiased. 

In Table~\ref{table:diff} we also compare the reported mass values of small planets with RV masses based on measurements done by two different instruments, namely HIRES and HARPS-N, and again report their difference HIRES$-$HARPS-N. This sample contains 4 planets including Kepler-10 c, which has measured masses discrepant between different instruments at the 3-$\sigma$ level \citep{2016Weiss,2016Petigura,2016Dai}.
Thus the differences between different RV instruments performing mass measurements is similar to the differences between RV and TTV measurements. 
The similar scale of discrepancy suggests that the small difference between RV and TTV observations of individual systems is insufficient to claim one of the methods is systematically biased compared to the other.

\begin{table}
\scriptsize
\caption{Masses of Planets Measured with Both RVs and TTVs}
\begin{tabular} { l | c c | c}
\hline
Planet & RV Mass  &  TTV Mass  & RV$-$TTV \\
&($M_\mathrm{Earth}$) & ($M_\mathrm{Earth}$) & \\
\hline
Kepler-18 b 	& $12\pm5$ \tablenotemark{a} 			& $18\pm9$ \tablenotemark{a}			& $-0.58\sigma$	\\
Kepler-18 c 	& $15\pm5$ \tablenotemark{a}			& $17.3\pm1.7$ \tablenotemark{a}		& $-0.44\sigma$	\\
Kepler-18 d 	& $28\pm7$ \tablenotemark{a}			& $8\pm1.3$ \tablenotemark{a}			& $+1.7\sigma$		\\
Kepler-89 d 	& $106\pm11$ \tablenotemark{b}		& $52.1^{+6.9}_{-7.1}$ \tablenotemark{c}	& $+4.2\sigma$		\\
Kepler-89 e 	& $35^{+18}_{-28}$ \tablenotemark{b}	& $13.0^{+2.5}_{-2.1}$ \tablenotemark{c}	& $+0.78\sigma$	\\
K2-19 b	 	& $31.8^{+6.7}_{-7.0}$ \tablenotemark{d}	& $44\pm12$ \tablenotemark{e}		& $-0.89\sigma$	\\
K2-19 c	 	& $26.5^{+9.8}_{-10.8}$ \tablenotemark{d}& $15.9^{+7.7}_{-2.8}$ \tablenotemark{e}	& $+0.80\sigma$	\\
Wasp-47 b	& $362\pm16$	\tablenotemark{f}	 	& $341^{+73}_{-55}$	 \tablenotemark{g}	& $+0.28\sigma$	\\
\hline 
\hline
Planet & HIRES RV Mass  & HARPS RV Mass & HIRES$-$HARPS \\
&($M_\mathrm{Earth}$) & ($M_\mathrm{Earth}$)  & \\
\hline
Kepler-10 b 	& $4.61\pm0.83$ \tablenotemark{h} 			& $3.30\pm0.49$ \tablenotemark{h}		& $+1.4\sigma$	\\
Kepler-10 c 	& $5.69^{+3.19}_{-2.90}$ \tablenotemark{h}	& $17.2\pm1.9$ \tablenotemark{h}		& $-3.1\sigma$	\\
K2-24 b	 	& $21.0\pm5.4$ \tablenotemark{i}			& $19.8^{+4.5}_{-4.4}$ \tablenotemark{j}	& $+0.17\sigma$	\\
K2-24 c	 	& $27.0\pm6.9$ \tablenotemark{i}			& $26.0^{+5.8}_{-6.1}$ \tablenotemark{j}	& $+0.11\sigma$	\\
\hline
\end{tabular}
\tablenotetext{1}{\cite{2011ApJS..197....7C}}
\tablenotetext{2}{\cite{2013Weiss}}
\tablenotetext{3}{\cite{2013ApJ...778..185M}}
\tablenotetext{4}{\cite{2016ApJ...823..115D}}
\tablenotetext{5}{\cite{2015MNRAS.454.4267B}}
\tablenotetext{6}{\cite{2012MNRAS.426..739H}}
\tablenotetext{7}{\cite{2015ApJ...812L..18B}}
\tablenotetext{8}{\cite{2016Weiss}}
\tablenotetext{9}{\cite{2016Petigura}}
\tablenotetext{10}{\cite{2016Dai}}
\label{table:diff}
\end{table}

\section{Effect of Planet Period}
\label{sec:period}

\subsection{Period differences}

A possible clue to the difference between the RV and TTV masses might be seen in the right panel of Figure~\ref{fig:mrpall}, which shows mass determination versus the planetary orbital period. One can see that on the average, the TTV (blue) periods are substantially longer than the RV (red) ones. As pointed out by S16, this difference could be the result of the dependence of the SNR of the two techniques on the orbital period. 

S16 showed that the SNR for an RV measurement is
\begin{equation}
\mathrm{SNR}_\mathrm{RV} \sim \frac{M_\mathrm{p}}{\sigma_\mathrm{RV} P^{1/3}}\ ,
\label{eqn:RVSNR}
\end{equation}
where $\sigma_\mathrm{RV}$ is the intrinsic uncertainty of a given RV measurement,
${M_\mathrm{p}}$ is the planetary mass
 and $P$ is the orbital period of the planet. For a TTV measurement S16 obtained
\begin{equation}
\mathrm{SNR}_\mathrm{TTV} \sim \frac{M_\mathrm{p} R_\mathrm{p}^{3/2} P^{5/6} }{\sigma_\mathrm{TTV} }\ ,
\label{eqn:TTVSNR}
\end{equation}
with $ R_\mathrm{p}$ the radius of the planet and $\sigma_\mathrm{TTV}$ the uncertainty of a point in the light curve (S16). 

We agree with these forms for an individual RV and TTV measurement, but point out that the vast majority of transiting exoplanets in the $<$8 $R_\mathrm{Earth}$ range of our interest have been discovered and characterized by the \Kepler mission, whose observing window was 4 years. Therefore, when fitting, e.g., a sine curve to the TTV measurements, shorter period planets have more data points in the fixed observational window \citep{2013Mazeh, 2016Holczer}. The number of transit (and thus TTV) observations, $N$, is $\propto$ $P^{-1}$ and, assuming statistical white noise properties, SNR $\propto$ $N^{1/2}$. Thus we suggest that Equation~\ref{eqn:TTVSNR} need be multiplied by a factor $P^{-1/2}$ to yield 
\begin{equation}
\mathrm{SNR}_\mathrm{KEPLER} 
\propto \frac{M_\mathrm{p} R_\mathrm{p}^{3/2} P^{1/3} }{\sigma_\mathrm{TTV} } \ .
\label{eqn:TTVSNRtrue}
\end{equation}

The SNR of an RV data set also increases as $N^{1/2}$ after the first orbital period (S16). However, unlike TTV measurements which relied nearly exclusively on \emph{Kepler}, there is not a clear practical limit to the length of time over which RV measurements may be made. Thus there is no period dependency on $N$, and the RV proportionality still holds when Equation~\ref{eqn:RVSNR} is multiplied by $N^{1/2}$ (excluding planets whose periods are so long that less than a few orbital periods have been observed). With or without our modification, it is obvious that it is easier for the RV technique to obtain a short-period solution, all other parameters being equal, while the TTV techniques can  obtain masses more easily for a longer period. The two relations can therefore account for the substantial period difference between the RV and TTV populations. 
A dependence of M--R relation  of the orbital period might therefore explain the mass difference between the two populations.

\subsection{M--R relation for different periods for the RV and the TTV masses}

\begin{figure}
\centerline{
\includegraphics[scale=0.3]{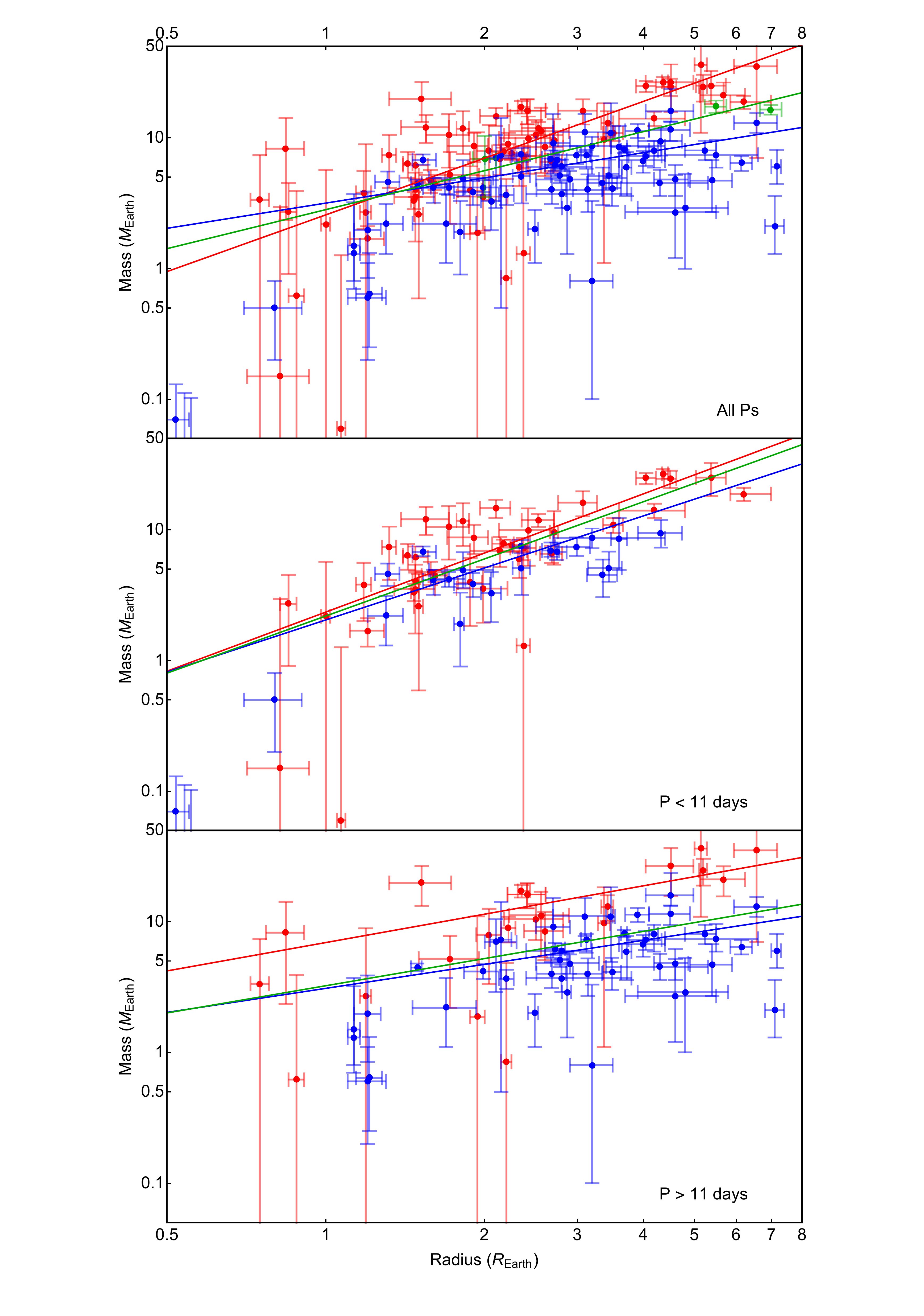}
}
\caption{Power-law fits to the masses and radii of planets measured via RVs (red), TTVs (blue), and both (green) including all data (top panel) and then broken into short and long period bins (bottom 2 panels). Note that the slopes of the power laws are consistent in each period bin, but a constant offset is remains in the long period bin. Parameters of the fits can be found in Table~\ref{table:plawfits}.
 }
\label{fig:plawfits}
\end{figure}

To study the mass-radius relationship and its dependence on observational technique, we fit all known masses and radii 
with a power-law M--R relation---$M_p\propto R_p^x$, where $M_p$ and $R_p$ are the planetary mass and radius. 
We then separated the data into RV and TTV measurements, and for each subsample we  performed a weighted total (or orthogonal) least squares linear fit \cite[e.g.,][]{1966CaJPh..44.1079Y,2007MeScT..18.3438K} to the logarithmic mass and radius with a Markov chain Monte Carlo search. The linear best fits for all measurements (green), RV measurements (red), and TTV measurements (blue) are shown as straight lines in the upper panel of Figure~\ref{fig:plawfits} and listed in the first line of Table~\ref{table:plawfits}.


\begin{table*}
\centering
\caption{Power Law Model Fits of the Form $M = 10^C \times R^E$}
\begin{tabular} { l | c  c | c  c | c c  }
\hline
Period Range &  \multicolumn{2}{|c|}{Combined Data Fit} &  \multicolumn{2}{|c|}{RV Fit} &  \multicolumn{2}{|c}{TTV Fit} \\
 				& Exponent 		& Constant 		& Exponent	 	& Constant 		 & Exponent		& Constant		\\
\hline 
All 				&$0.99\pm0.04$ 	& $0.45\pm0.02$	& $1.45\pm0.07$	& $0.41\pm0.03$	& $0.64\pm0.06$	& $0.50\pm0.03$	\\
$P<11$ days 	 	&$1.46\pm0.06$ 	& $0.34\pm0.03$	& $1.50\pm0.08$	& $0.36\pm0.03$	& $1.34\pm0.19$	& $0.31\pm0.07$	 \\
$P>11$ days 		&$0.69\pm0.06$ 	& $0.51\pm0.03$	& $0.73\pm0.19$	& $0.83\pm0.09$	& $0.61\pm0.06$	& $0.49\pm0.03$	 \\
\end{tabular}
\label{table:plawfits}
\end{table*}

The upper panel of Figure~\ref{fig:plawfits} shows the linear logarithmic regressions obtained separately for the TTV and the RV masses using planets of all orbital periods. The linear fit for the RV mass measurements, with exponent of $1.45\pm 0.07$,  is substantially steeper than that of the TTV measurements, with exponent of $0.64\pm 0.06$. The larger exponent could suggest that masses measured via the RV technique increase with radius significantly more quickly than the masses obtained by the TTV measurements. 

To see if this is true, and based on the previous subsection, we divided the measurements into short- ($P<11$ day) and long-period ($P>11$ day) ranges. This period boundary was chosen empirically by identifying the maximum period for which planets with lower periods showed  a similar M-R relationship. Figure~\ref{fig:plawfits} and the two lower lines of Table~\ref{table:plawfits} reveal a clear difference in planet properties in each period bin. In each of these bins we again fitted linear regressions to the RV and TTV masses, and for these two data sets combined. Three surprising features emerged. First, the difference between the exponents of the M--R relation obtained for the TTVs and the RVs at the same period range disappeared. Second, there is a significant difference between the logarithmic slope of the M--R relation in the short-period range, $1.46\pm 0.06$,  and the long-period range, of  $0.69\pm 0.06$. These values are similar to the slopes obtained by the RV and TTV measurements individually over all period ranges, suggesting the different masses obtained by RVs and TTVs are probably a result of the different period range observed. As previously noted, this observational difference is due to the different period dependency of the methods' sensitivities (see Equations~\ref{eqn:RVSNR} and~\ref{eqn:TTVSNRtrue}). 
Additionally, none of the measured M-R relationships agree with the frequently used power-law slope of 2.06 inspired by Solar System objects \citep{2011Lissauerb}.

A third feature that emerged in Figure~\ref{fig:plawfits} lower panel 
is a statistically significant offset between the RV and TTV measurements in the long-period bin. Almost all the masses derived by RVs are higher than those derived by TTVs. In fact, there is no significant RV mass determination below 
$3$--$4 M_\mathrm{Earth}$. We attribute this difference to the fact that TTVs readily can detect long-period planets of low mass, whereas RV measurements are decreasingly sensitive to low-mass planets at longer orbital periods (see Equation~\ref{eqn:RVSNR}). This decline in sensitivity could lead to a higher-mass threshold for the RV detection and the apparent observed mass discrepancy. Additionally, when the stellar rotation period ($\sim$10--50 days for main sequence FGKM stars) approaches the planetary orbital period, the RV amplitude detection threshold can worsen due to challenges distinguishing between star spot activity and the planet signal. This degeneracy may prevent low-mass planets with long periods from being confidently detected by the RV technique even with a large amount of observational data. 

\section{Planet Period Ratios}
\label{sec:prats}

\begin{figure*}
\centerline{
\includegraphics[width=0.49\textwidth]{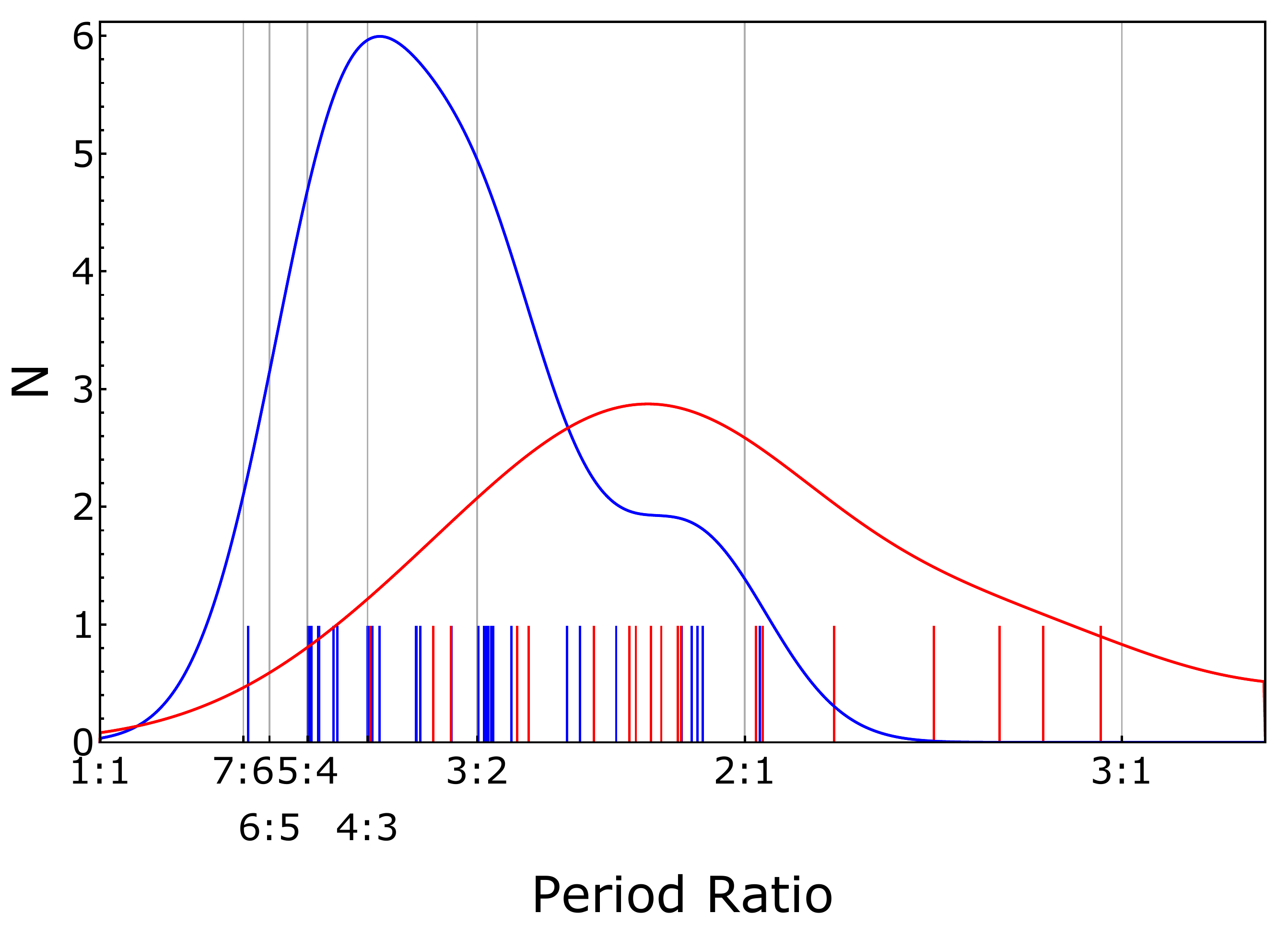}
\includegraphics[width=0.5\textwidth]{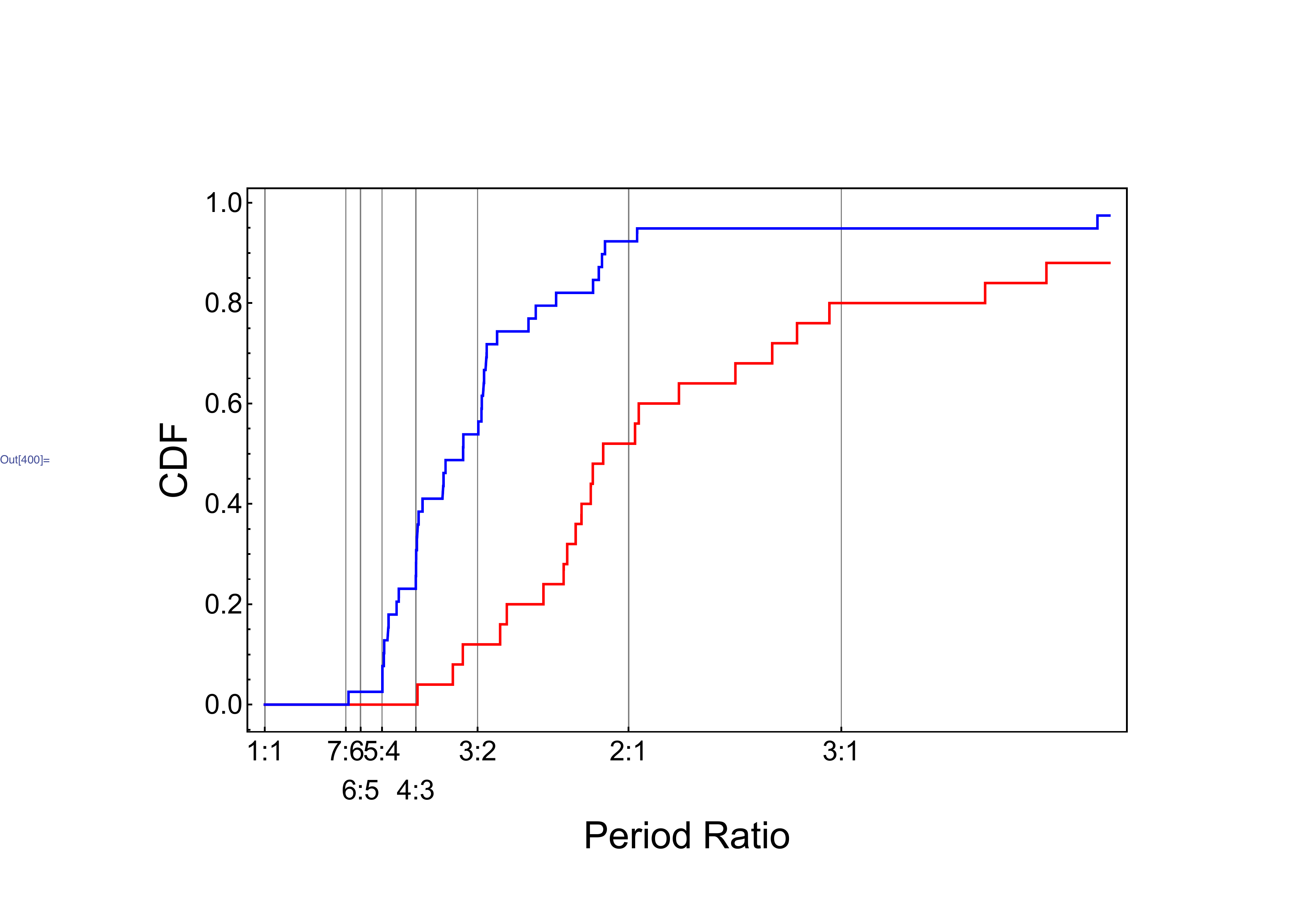}
}
\caption{ \emph{Left:} The period ratios of pairs of planets with masses measured via TTV (blue) and RV (red) are marked as ticks on the horizontal axis with logarithmic scaling. A broad Gaussian kernel is used to produced a smooth histogram of both distributions. Note that many RV planets have no known companions and are therefore not included on this plot. \emph{Right:} The cumulative distribution function (CDF) of the TTV and RV planet period ratios. There are a few very-widely spaced RV planets, which causes the RV (red) CDF to reach 1.0 beyond the limit plotted in this panel. 
  }
\label{fig:prats}
\end{figure*}

The SNR of TTV signals depends strongly on the period ratio of the perturbing adjacent planet. The signal is greatest for compact systems near low-order mean motion resonances \citep[e.g.,][Figure~8]{2005Agol,2012Lithwick,2016Hadden}, all else equal. On the other hand, RV mass detection of a planet is potentially hampered by the existence of another planet with a similar period. As a result, there is an inherent observational bias for more tightly-spaced  planetary systems measured with TTVs compared to RVs, as  shown in Figure~\ref{fig:prats}. 
A two-sample Kolmogorov-Smirnoff test returns $\alpha=4\times10^{-6}$, indicating a clear statistical difference between the two distributions.

\section{Summary}
\label{sec:discussion}

Following S16, we have shown that the difference between RV and TTV masses can be attributed to the different SNR dependence on orbital period between the two techniques.  This difference causes a substantial difference in the period range of the two techniques---RV masses are mostly obtained for relatively short orbital periods, while the TTVs masses are weighted towards longer periods. When we divide the data into short- and long-period orbits, most of the differences between the masses of the two techniques disappear. This suggests that the underlying mass distribution measured by the RV and TTV techniques has similar properties. 

We have found that the exponent in  the  power-law relation 
$M_p\propto R_p^x$ that presumably characterizes the M--R relation is substantially different for the short and the long orbital periods---the exponent best value is
$1.46\pm 0.06$ for the short-period planets and $0.69\pm 0.06$ for the long-period range. 
Both values suggest that, on average, the density of the planets (proportional to $M_p/R_p^3$), is decreasing as a function of the planetary radius, as expected from considerations of planetary composition. However, our analysis suggests  that the rate of density decrease depends on the orbital period. If confirmed, this has to be accounted for by planetary formation or evolution theories.
We also demonstrate the expected more compact orbital period ratios of planets measured via TTVs compared to RVs.

\section*{Acknowledgements}
We thank Jason Steffen for his seminal paper which inspired this study and for helpful discussions, and we thank Daniel Fabrycky, Jack Lissauer, William Welsh, and an anonymous reviewer for insightful comments which considerably improved this work. We greatly appreciate Daniel Jontof-Hutter for providing us with a tabulation of small exoplanet data through 2015 and illuminating comments. We also thank Josh Burkhart for making his Mathematica MCMC code publicly available (https://github.com/joshburkart/mathematica-mcmc). This research has made use of NASA's Astrophysics Data System. This research has received funding from the Israeli Centers for Research Excellence (I-CORE, grant No.~1829/12).


\clearpage

\end{document}